\documentclass[12pt]{iopart}

\usepackage{iopams} 
\usepackage{graphicx}
\usepackage{epsfig}

\begin{document}

\title[Consensus formation on adaptive networks]{Consensus 
formation on coevolving networks: 
groups' formation and structure}

\author{Balazs Kozma$^1$ and Alain Barrat$^{1,2}$}
\address{$^1$LPT, CNRS, UMR 8627, and
Univ Paris-Sud, Orsay, F-91405, France}
\address{$^2$Complex Networks Lagrange Laboratory, ISI Foundation,
Turin, Italy}

\ead{kozmab@th.u-psud.fr, alain.barrat@u-psud.fr}
\begin{abstract}

We study the effect of adaptivity on a social model
of opinion dynamics and consensus formation.  We analyze how the
adaptivity of the network of contacts between agents to the underlying
social dynamics affects the size and topological properties of groups
and the convergence time to the stable final state. We find that,
while on static networks these properties are determined by
percolation phenomena, on adaptive networks the rewiring process leads
to different behaviors: Adaptive rewiring fosters group formation by
enhancing communication between agents of similar opinion, though it
also makes possible the division of clusters. We show how the
convergence time is determined by the characteristic time of link
rearrangement. We finally investigate how the adaptivity yields
nontrivial correlations between the internal topology and the size of
the groups of agreeing agents.

\end{abstract}

\submitto{\JPA}

\section{Introduction}

In the last years, many efforts have been devoted to the understanding
of how the behavior of models of interacting agents is affected by the
topology of these interactions. Various simple models have indeed been
defined in order to study how local rules of evolution can lead to the
emergence of global phenomena. This topic, at the center of the
statistical physics, has been connected to the field of social
sciences, leading to the so-called sociophysics
\cite{sociophysics}. The statistical physics approach starts from the
simplest possible models, in which each agent is defined by an
internal state, which evolves according to certain rules of
interactions with the neighboring agents. The internal state can be an
Ising variable taking only two possible values
\cite{Galam:1982,Krapivsky:1992,Krapivsky:2003}, a vector of traits
\cite{Axelrod:1997}, a continuous variable
\cite{Deffuant:2000,BenNaim:2003,StaufferOrtmanns2003,BenNaim:2005,Amblard:2004},
or have a more complex structure
\cite{Steels:1996,Baronchelli:2006,Dallasta:2006c}. The {\em local}
interaction rules may involve the simple imitation of a neighbor, an
alignment to a local majority, or more involved negotiation processes.
The main question concerns in all cases the possibility of the
appearance of a global consensus, defined by the fact that all agents'
internal states coincide, without the need for any central authority
to supervise such behavior. Alternatively, polarized states in which
several coexisting states or opinions survive can be obtained, and
the parameters of the model drive the transition between consensus
and polarization.

The science of complex networks has recently led to an intense activity
devoted to the characterization and the understanding of the topology of
man-made and natural networks
\cite{Granovetter:1973,Wasserman:1994,Albert:2002,Dorogovtsev:2002a,Dorogovtsev:2003a,Newman:2003b,pastorsatorras2004,Gonzalez:2006},
such as social networks, which determine how social agents interact.
Numerous studies have revealed the ubiquity of various striking
characteristics, such as the small-world effect: although each node has a
number of neighbors which remains small with respect to the total number of
agents, only a small number of hops suffices to go from any agent to any
other on the network.  This has prompted the investigation of the effect of
various interaction topologies on the behavior of agents connected
according to these topologies, highlighting the relevance of small-world
and heterogeneous structures (see for example
\cite{Castellano:2005,Sood:2005,Suchecki:2005a,Suchecki:2005b}).

More recently, the focus has shifted to take into account the fact
that many networks, and in particular social networks, are dynamical
in nature: links appear and disappear continuously in time, on many
timescales. Moreover, such modifications of the network's topology do
not occur independently from the agents' states but as a feedback
effect: the topology determines the evolution of the agents' opinions,
which in its turn determines how the topology can be modified
\cite{Zimmermann:2004,Ehrhardt:2006,Gil:2006,Holme:2006,Caldarelli:2006,Stauffer:2006,Kozma:2007};
the network becomes adaptive.  In this framework, we consider here the
coevolution of an adaptive network of interacting agents, and
investigate how the final state of the system depends on this
coevolution. We focus on the Deffuant model in which opinions are
continuous variables and neighboring agents which have close enough
opinions (as determined by a tolerance parameter) can reach a local
consensus. In our model, the rate of evolution of the network's
topology is tunable and represents one of the parameters. We focus on
simple evolution rules of the topology, that do not require prior
knowledge of the state of agents to which new links are
established. We study the role of the various parameters such as the
tolerance of agents and the rate of topology evolution. As already
shown in a previous investigation \cite{Kozma:2007}, the possibility
of the interaction network to adapt to the changes in the opinion of
the agents has important consequences on the evolution mechanisms and
on the structure of the system's final state. We focus here on the one
hand on the convergence time needed to reach this final state, and on
the other hand on the detailed characterization of this state, in
which various groups of agents are formed, each group with a distinct
uniform opinion.

The paper is organized as follows. We define precisely the model
in section \ref{sec:model}. We briefly recall in section
\ref{sec:transitions} the results obtained in \cite{Kozma:2007}
concerning the comparison of the Deffuant model on static and adaptive
networks, and in particular how the rewiring affects the transitions
between consensus and polarized states.  We focus in section
\ref{sec:tc} on the issue of convergence time, i.e. on how the
parameters determine the time to reach a global or partial consensus.
Finally, we consider in section \ref{sec:clusters} the detailed
structure of the final state of the agents. We investigate in
particular how the size of the various connected components, or groups
of agents, is linked with their average number of connections.

\section{The model}
\label{sec:model}

We consider the Deffuant model for the evolution of opinions, in which $N$
agents are endowed with a continuous opinion $o \in [0:1]$
\cite{Deffuant:2000,BenNaim:2003,StaufferOrtmanns2003,BenNaim:2005,Amblard:2004}. Starting
from random values, the agents' opinions evolve through binary interactions
according to the following rules: at each timestep $t$, two neighbouring
agents are chosen at random. If their opinions are close enough, i.e., if
$|o(i,t)-o(j,t)|\le d$, where $d$ defines the tolerance range or threshold,
they can communicate, and the interaction tends to bring them closer,
according to the rule
\begin{eqnarray}
\nonumber
   o(i,t+1)=o(i,t)+\mu(o(j,t)-o(i,t)) \\
   o(j,t+1)=o(j,t)-\mu(o(j,t)-o(i,t))
\label{deffuant}
\end{eqnarray}
where $\mu \in [0,1/2]$ is a convergence parameter. For the sake of
simplicity and to avoid a too large number of parameters, we will
consider the case of $\mu=1/2$: $i$ and $j$ adopt the same
intermediate opinion after communication \cite{BenNaim:2003}. The
tolerance parameter plays a crucial role in the ability of the
population of agents to reach a global consensus or not. It is indeed
intuitively clear that, for large tolerance values, agents can easily
communicate and converge to a global consensus. On the contrary, small
values of $d$ naturally lead to the final coexistence of several
remaining opinions
\cite{Deffuant:2000,BenNaim:2003}.

For large populations, it may be more realistic to consider that the
interactions between agents define a network with a finite average
connectivity: each agent has a limited number of neighbours and can
not a priori communicate with all the other agents. A typical example
of such an interaction network structure is given by an uncorrelated
random graph in which agents have $\bar{k}$ acquaintances on average,
i.e. the initial network corresponds to an Erd\H{o}s-R\'{e}nyi network
with average degree $\bar{k}$. While such a topology lacks many
interesting features displayed by real social networks, such as degree
heterogeneity or community structures, it is nevertheless interesting
to first consider such a simple case as a reference frame. Moreover,
we focus in this work on another important aspect of real networks:
the fact that their topology may evolve on the same timescale as the
agents' opinions. Agents can indeed break a connection or establish new
ones, depending on the success of the corresponding relationship. The
rules defining the evolution of the network topology can be modeled in
many different ways. A possibility is to consider that links decay at
a constant rate, independently from the agents' opinions
\cite{Ehrhardt:2006}. In the case of opinion dynamics, we consider
instead that only neighbouring agents with far apart opinions (i.e.,
$|o(i,t)-o(j,t)| > d$) may terminate their relationship
\cite{Kozma:2007}. In order to keep the average
number of interactions constant, a new link is then introduced between
one of the agents having lost a connection and another agent, chosen
at random. The new link may of course break again if the newly
connected agents have too-far-apart opinions. The rewiring process
thus occurs as a random search for agents with close-enough opinions.

Even a simple rewiring process such as the one depicted above leads to
the introduction of two new parameters. The first one, $w$, quantifies
the relative frequencies of the two following processes: a local
opinion convergence for agents whose opinions are within the tolerance
range, and a rewiring process for agents whose opinions differ more.
At each time step $t$, a node $i$ and one of its neighbors $j$ are
chosen at random. With probability $w$, an attempt to break the
connection between $i$ and $j$ is made: if $|o(i,t)-o(j,t)| > d$, the
link $(i,j)$ is removed and a new link is created. With probability
$1-w$ on the other hand, the opinions evolve according to
(\ref{deffuant}) if they are within the tolerance range. The second
parameter concerns the creation of a new link whenever a link $(i,j)$
has been removed: a new node $k$ is then chosen at random, and with
probability $p \in [0:1]$ a link $(i,k)$ is created, while with
probability $1-p$ the new link instead connects $j$ and
$k$. Since $j$ is chosen as a neighbour of a randomly chosen node $i$,
it will have on average a larger degree. Larger $p$ therefore favors
the removal of links from larger degree nodes, while small $p$ means
that large degree nodes preferentially keep their links. We will see
in section \ref{sec:clusters} how the parameter $p$ affects the final
structure of the agents groups.

Both for static and adaptive networks, the system is initialized in a state
where the agents have a random opinion between $0$ and $1$ and the
network configuration is Erd\H{o}s-R\'{e}nyi. The averages were generated
over $10$ to $100$ different networks, as specified in the captions of the
figures. The dynamics stops when no update is anymore possible. If $w>0$,
this corresponds to a state in which no link connects nodes with different
opinions. This can correspond either to a single connected network in which
all agents share the same opinion, or to several disconnected clusters
representing different opinions. For $w=0$ on the other hand, the final
state is reached when neighboring agents either share the same opinion or
differ of more than the tolerance $d$.

\section{Static versus adaptive}
\label{sec:transitions}

\begin{figure}[thb]
\vspace*{0.65cm}
\centering
\includegraphics[width=.4\textwidth]{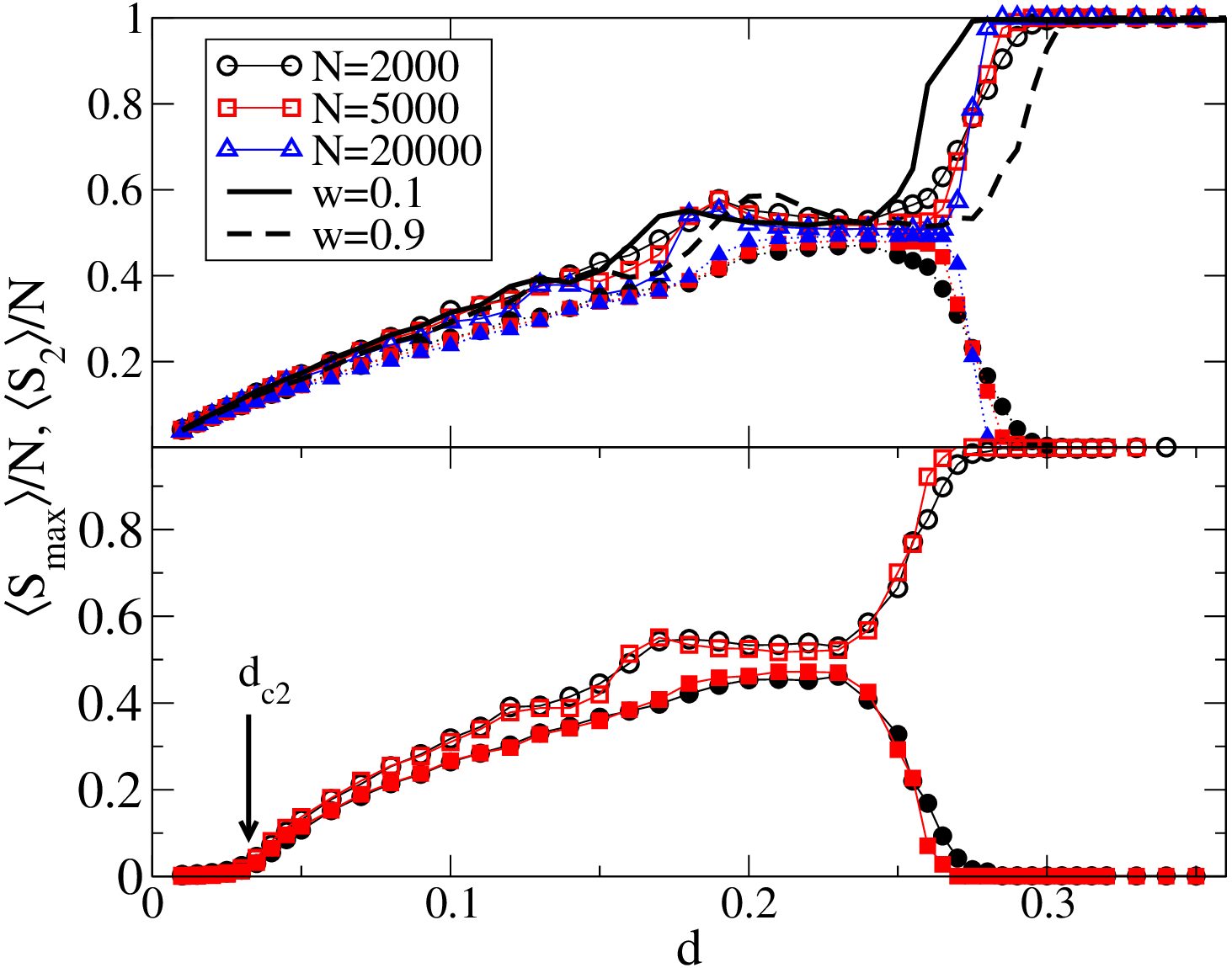}
\hspace{.3cm}
\includegraphics[width=.47\textwidth]{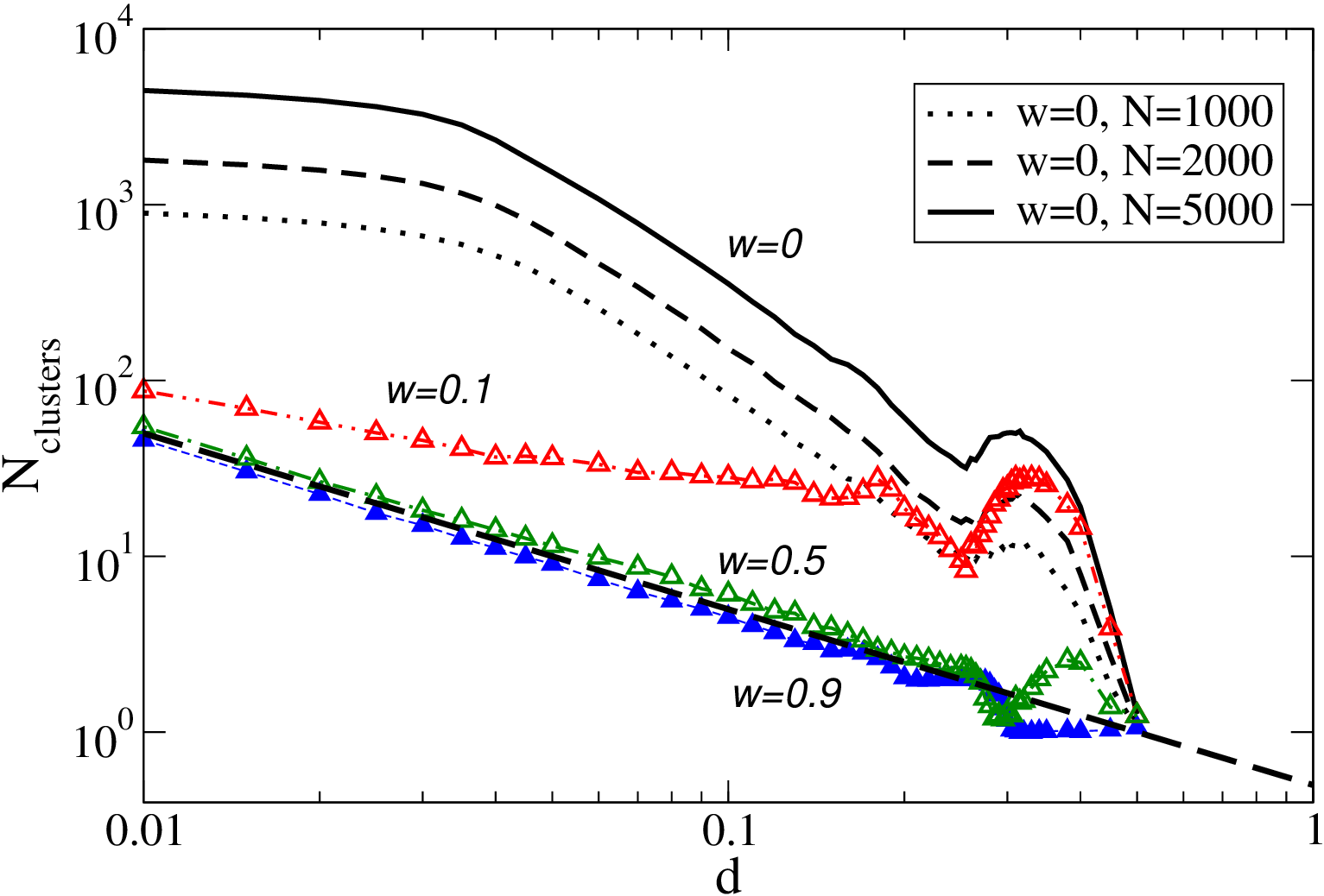}
\vspace*{-0.0cm}
\caption{
\label{fig:sizes}(Color online) Left figure:
Size of the largest (empty symbols) and second-largest (filled
symbols) clusters in the final state, as a function of the tolerance
value on a static Erd\H{o}s-R\'enyi network with average degree
$\bar{k}=10$ for different system sizes (bottom) and on adaptive
networks with the same average degree and $w=0.5$. Continuous and
dashed curves give the size of the largest cluster for $N=5000$ and
$w=0.1$ and $w=0.9$, respectively: the consensus-to-polarized
transition point is shifted to larger and larger tolerance values as
the adaptivity of the network increases.\\
Right figure: Number of clusters in the final state
as a function of the tolerance of the agents for
different system sizes on a static Erd\H{o}s-R\'enyi
network with $\bar{k}=10$, and on 
adaptive networks with the same average degree, $N=5000$, and
different rewiring rates. The dashed line corresponds to the value
$1/(2d)$, asymptotically valid at small tolerances in the mean-field case.
Averages were generated over 100 different networks.}
\end{figure}

In this section, we focus on the case $p=0$ and compare the results of
the opinions evolution on a static and a dynamically adaptive network.
Figure \ref{fig:sizes} displays for both cases the size of the largest
($\langle S_{max}\rangle/N$) and second largest ($\langle
S_2\rangle/N$)) opinion clusters in the final state, where a cluster
is defined as a connected group of agents sharing the same opinion. In
both cases, at large tolerance $d$, a global consensus is achieved,
with a single cluster containing all agents. A jump of $\langle
S_{max}\rangle/N$ from a value close to $1$ to a value close to $1/2$
is observed at a critical value $d_{c1}(w)$. Interestingly, and as
hypothesized in \cite{Kozma:2007}, the structure of the largest
cluster changes as $d$ decreases towards $d_{c1}(w)$: the average
shortest path between two nodes increases as $d \to d_{c1}(w)$ (data
not shown), because the cluster typically acquires a community structure where
only a few nodes or links keep the communities connected. Moreover,
the critical value $d_{c1}(w)$ increases with $w$ (see
Fig. \ref{fig:sizes}). If the rewiring is more frequent, it allows
more easily to break the network in two pieces, since agents can more
easily search for other agents with whom they can communicate, and
break ties with the ones with too different opinions: the formation of
different clusters is favored, and larger tolerance values are
necessary to achieve consensus.

At $d < d_{c1}$, an apparently polarized state is entered, with a
first and second-largest clusters of similar extensive sizes, and
apparently similar behavior for static and adaptive networks. A
difference appears however at small tolerance values: for static
networks, $\langle S_{max}\rangle/N$ vanishes for $d < d_{c2}$ in the
thermodynamic limit. The final state of the system is then {\em
fragmented}, with no cluster of extensive size.  This
polarized-fragmented transition is due to the finite connectivity of
the agents and corresponds to a percolation phenomenon. An agent $i$
with $k$ connections and tolerance range $d$ will indeed have on
average $2dk$ neighbours with whom to communicate. For an average
degree $\bar{k} < 1/(2d)$, there will not be any percolating paths of
agents with close enough opinions and only very small clusters of
agreeing agents can be formed.  In the case of adaptive networks on
the other hand, the fragmented phase disappears as soon as the
rewiring of the links is enabled. The size of the largest component
decays smoothly but remains extensive as the tolerance
decreases. Rewiring processes thus allow the small clusters, initially
non-percolating, to create links between them and reach
extensive sizes even below the polarized-to-fragmented transition
appearing on static networks.

Further insight into the differences between static and adaptive
networks is provided by the number of opinion clusters in the final
state, $N_{clusters}$, shown in Fig. \ref{fig:sizes}. For $d_{c2} \le
d \le d_{c1}$, an extensive number of clusters is indeed obtained in
the static case, saturating at ${\cal O}(N)$ at $d_{c2}$. The system
presents therefore a {\em ``false''-polarized} state, with a
coexistence of macroscopic opinion clusters with an extensive number
of finite size clusters. As $d$ decreases, more and more macroscopic
clusters appear, as in mean-field \cite{BenNaim:2003}, but there is
also an extensive proliferation of finite size ``microscopic''
clusters.  For adaptive networks, the number of clusters is much
smaller, and decreases as $w$ increases. In fact, the precise
investigation of the cluster size distribution reveals that the
density of nodes in non-extensive clusters vanishes in the
thermodynamic limit \cite{Kozma:2007}.  The polarized phase on
adaptive networks differs therefore strongly from the one on static
networks: thanks to the possibility of link rewiring, agents who would
remain isolated (or in very small groups) on a static network may
manage to find agents with whom to communicate and thus enter a
macroscopic cluster. Without rewiring on the other hand, a
macroscopic number of agents remain in fragmented components which
coexist with few macroscopic clusters.

\section{Convergence time}
\label{sec:tc}

\begin{figure}[thb]
\vspace*{0.65cm}
\centering
\includegraphics[width=.475\textwidth]{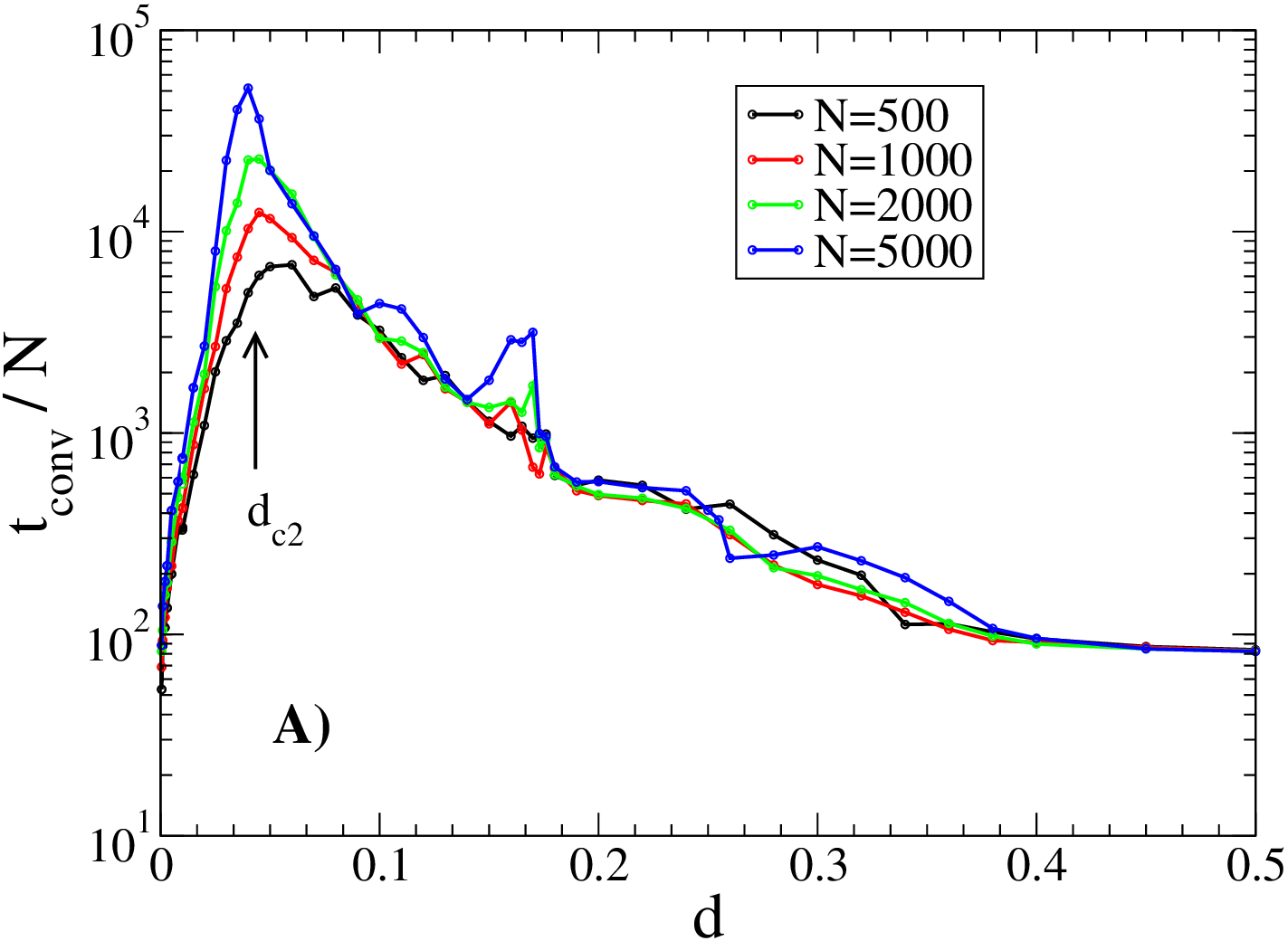}
\hspace{.3cm}
\includegraphics[width=.445\textwidth]{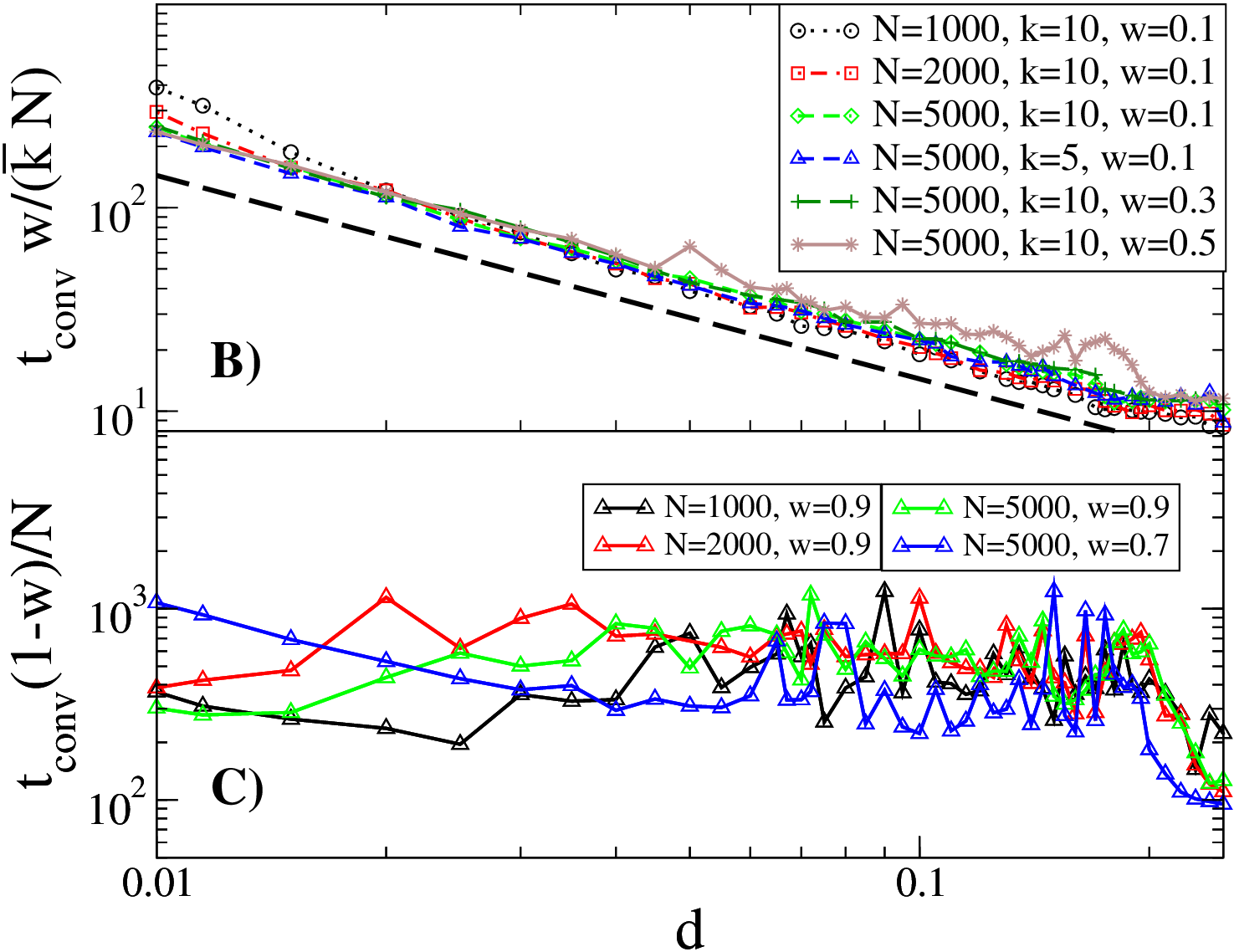}
\vspace*{-0.0cm}
\caption{
\label{fig:tConv}(Color online) A) convergence time, measured in the number
of simulation steps, on static networks for different system sizes as a
function of the tolerance, $d$, when $\bar{k}=10$ (The data points were
generated averaging over $10$ to $100$ realizations). B) rescaled
convergence time on adaptive networks for small rewiring rates as a
function of $d$ for different system sizes, average degrees, and rewiring
rates (Data points averaged over $30$ realizations.). The dashed line is
proportional to $1/d$. C) Rescaled convergence time on adaptive networks as
a function of $d$ for different system sizes and rewiring rates
($\bar{k}=10$) averaged over $100$ realizations.  }
\end{figure}

On static networks, the time to converge to the final state of the
system, $t_{conv}$, is determined by the topological properties of the
opinion clusters. In turn, the behavior of this topology is mostly
determined by the distance of the tolerance of the agents from that of
the polarized-to-fragmented transition, $d_{c2}$ (see
Fig. \ref{fig:tConv}): For $d_{c2}<d$, $t_{conv}$ grows linearly with
the system size and increases as $d$ decreases: the clusters formed by
agents who can communicate become more and more tree-like, which slows
down the convergence to a common opinion. As $d \to d_{c2}$,
$t_{conv}$ diverges as a signature of the phase transition.  For
$d<d_{c2}$, the clusters of agents with close enough opinions become
small and the convergence time decreases as $d$ decreases.

On adaptive networks, two scenarios are possible: on the one hand, if the
network evolution is slow compared to the timescale of opinion
formation, $t_{conv}$ is mostly determined by the characteristic time
of topological cluster formation, $t_l$ \footnote{It is important to
note that, even in this limit, the network topology cannot be
considered static and adaptivity plays an important role in the early
time evolution of the system too, as shown in \cite{Kozma:2007}.}. The
scaling of $t_l$ and therefore $t_{conv}$ can be estimated by
considering a typical opinion cluster: its size is proportional to the
tolerance range of the agents, $2d$; the number of its links which
need to be rewired is proportional to the total number of links
($\propto \bar{k} N$), and to the amount of opinions outside of the
tolerance range ($\propto (1-2d)$); the probability to rewire a link
towards an agent with a close enough opinion is moreover $\propto d$
and the time between two link updates is $\propto 1/w$ so that
\begin{equation}
t_l \propto \bar{k}(1-2d)/(w d) .
\label{tl}
\end{equation}
Figure \ref{fig:tConv}B) shows the rescaled convergence time
$w t_{conv}/\bar{k}$ as a function of $d$ for different parameter 
values, in good agreement with Eq. (\ref{tl}).
If the network evolution is fast, on the other hand, it is possible
for an agent to rewire most links towards agents with close
opinions in a short time. This scenario takes place when $t_l$ is less
than or comparable to $1/(1-w)$, the average time between two
fruitful discussions. In this case, the convergence time is expected
to scale simply as $1/(1-w)$, as indeed shown in 
Fig. \ref{fig:tConv}C).

\section{Group structure}
\label{sec:clusters}

\begin{figure}[thb]
\vspace*{0.65cm}
\centering
\includegraphics[width=.47\textwidth]{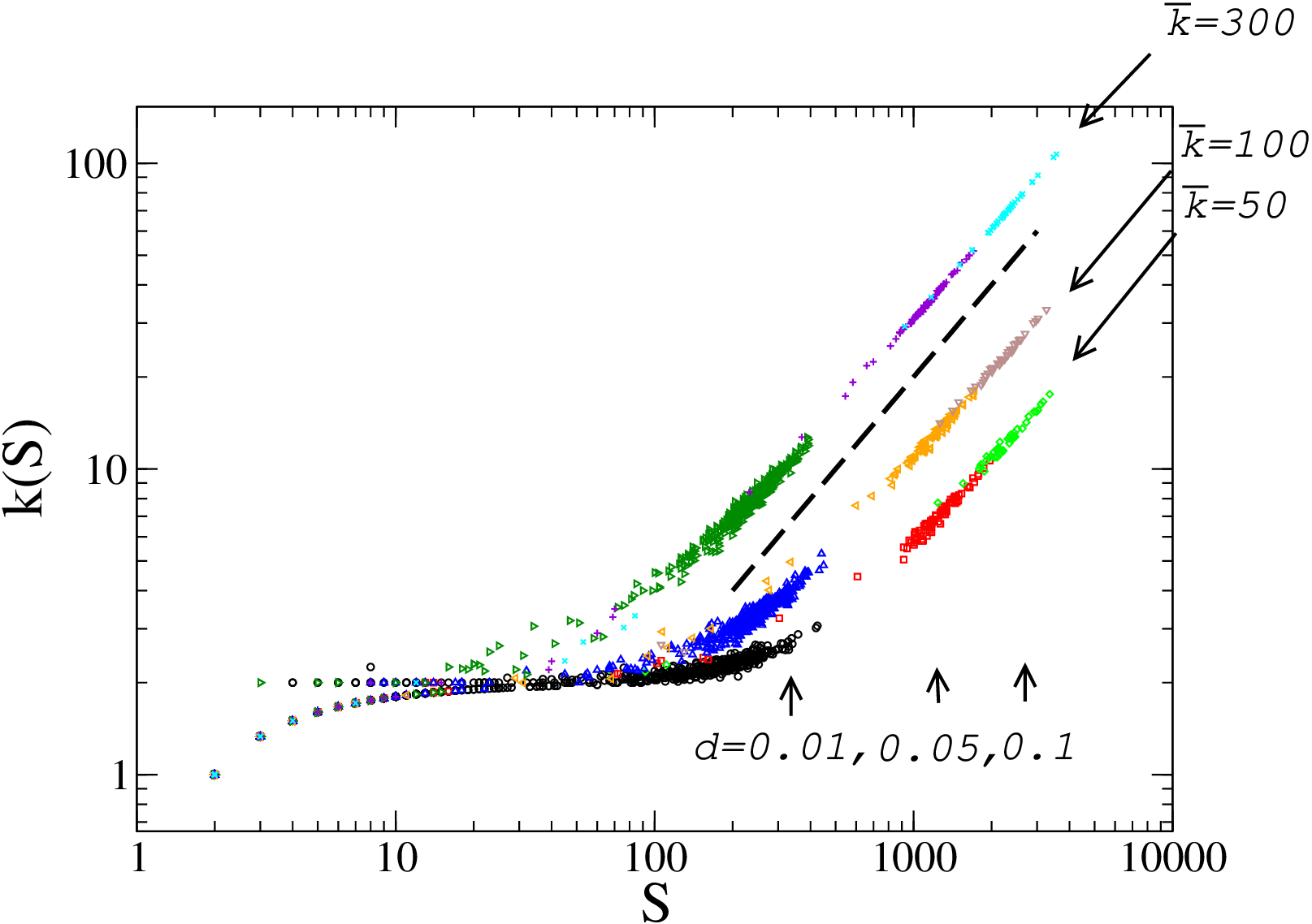}
\hspace{.3cm}
\includegraphics[width=.47\textwidth]{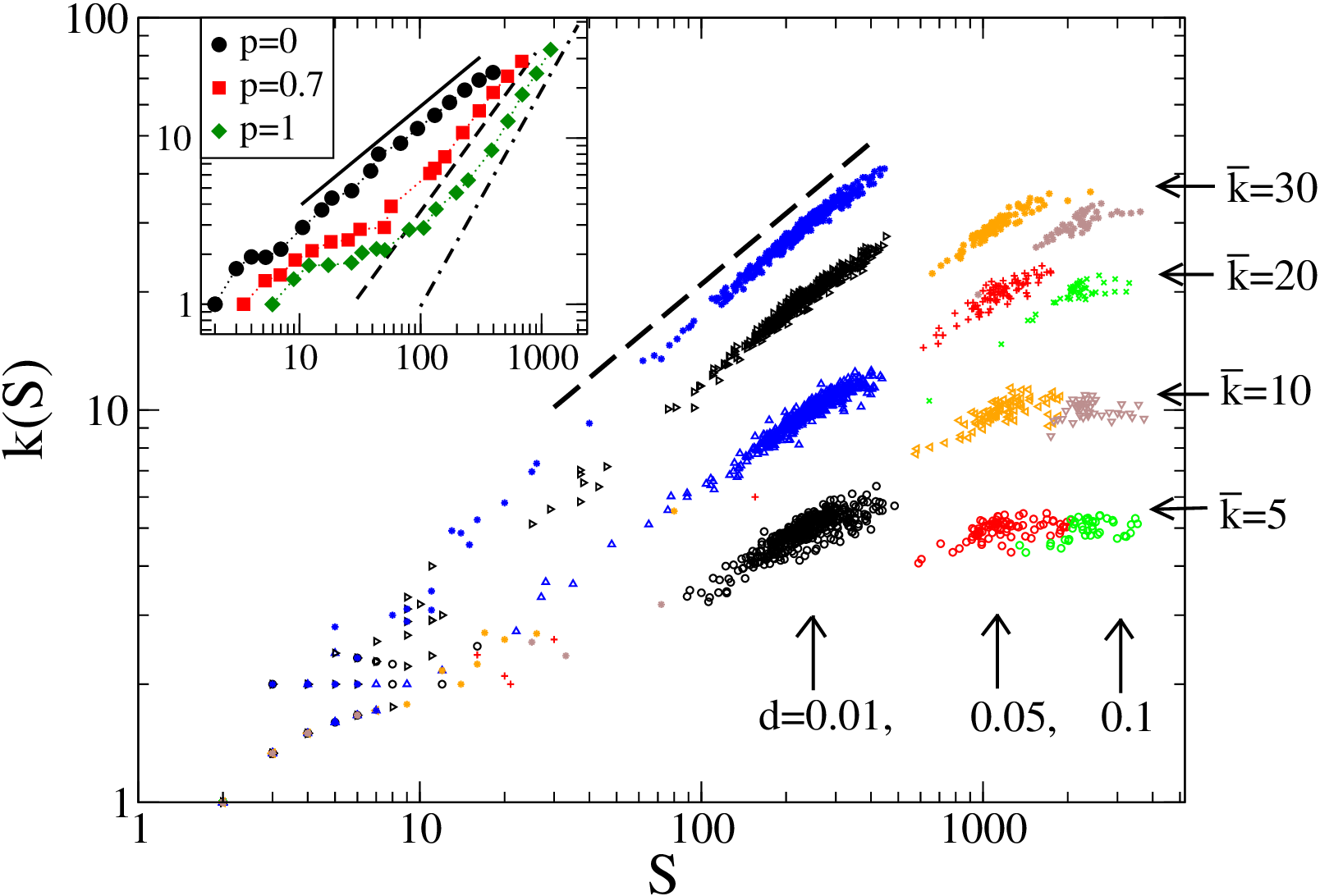}
\vspace*{-0.0cm}
\caption{
\label{fig:k(s)}(Color online) Left figure: average degree of the agents
sharing the same opinion as a function of the size of the uni-opinion
groups, $S$, on static networks. The dashed line corresponds to a linear
behaviour. Right figure: same in the case of adaptive networks when $w=0.5$
and $p=0$ for different average degrees and tolerance values.  The dashed
line shows a power-law $S^{0.6}$. Inset: the binned average of the same
measurement for $\bar{k}=20$, $d=0.01$, $w=0.5$, and different values of
$p$.  The lines correspond to the power laws $S^{0.6}$, $S$, and $S^{1.3}$.
In both figures, $N=10^4$ and data points were collected from $10$
realizations for each set of parameter values.}
\end{figure}

Once the population of agents has reached its final state, an
interesting question concerns the structural differences between the
various opinion groups that have been formed. The most basic property
one can investigate is the average degree of an agent. It turns out
that the average degree of an agent inside a group is strongly
correlated with the size of the group.  On static networks, the
average degree of a cluster is a linear function of its size (left
plot of Fig. \ref{fig:k(s)}), which can be explained as follows: for a cluster
of size $S$, the probability for a node to have a link pointing
towards this cluster is simply $S/N$, and a node of degree $k$ will 
have on average $kS/N$ links pointing towards other nodes in the cluster.
The average ``in-degree'' of the nodes in a cluster of size $S$ is therefore
$S\bar{k}/N$ (assuming that there is no correlation between the degree
of a node in the network and the cluster to which it finally belongs)
\footnote{Moreover, the average degree cannot be less than two, except for
starlike clusters where the average degree can be between $1$ and $2$, which
explains the flattening of the curves at small cluster sizes.}.

On adaptive networks on the other hand, the linear relationship is no longer
valid, as shown in Fig. \ref{fig:k(s)}. At small tolerance values, many
clusters are obtained, with very different sizes. A power-law-like
relationship appears between the clusters' size and degree. The behaviour
depends on $p$, the parameter of the rewiring rule: a sublinear relationship
holds for small $p$ values while a superlinear behaviour appears for $p$ close
to $1$.  Since these cases correspond to the situations when the development
of the opinion clusters takes place at a much shorter timescale than that of
the topological clusters \cite{Kozma:2007}, an analytical treatment of the
problem is possible, investigating the diffusion of the links between the
clusters of constant opinions at a mean-field level \cite{AGandBK}. For
large tolerance values, the cluster's average degree becomes less correlated
with its size (right graph of Fig. \ref{fig:k(s)}): this is due to the fact
that the clusters correspond to large fractions of the original network and
their average degree saturates at $\bar{k}$.

\section{Conclusions}

In this work we have investigated how adaptive network topology can influence
the final state and structure of the Deffuant model compared to its behavior
on static Erd\H{o}s-R\'enyi networks. While on static networks the model
exhibits three different phases, the fragmented phase, present at very small
tolerance values and characterised by the lack of extensive-size clusters,
disappears on adaptive networks since rewiring allows small groups to connect
to each other and grow to macroscopic sizes for all tolerance values. The
consensus-to-fragmented transition is shifted to larger tolerance values on
adaptive networks since rewiring promotes the division of agents with too
different opinions. We have found that the convergence time of the system on
static networks is determined by the distance of the tolerance of the agents
from that at the polarized-to-fragmented transition, $d_{c2}$. On
adaptive networks instead, 
the convergence time is set either by the time it takes to
successfully rewire the links between disagreeing agents or simply by the
frequency of opinion updates, depending on which of these timescales is larger
than the other.  Finally, probing the local structure of the groups by
measuring the average degree of them reveals nontrivial correlations 
between the size of a group and its average degree: The average degree is
either a sub or superlinear function of the size, determined by the parameter
of the rewiring rule, $p$. Adaptive dynamics moreover display robust features
with respect to small changes in the opinion-update rule contrarily to what
happens on static networks, on which for example the consensus-to-polarized
transition disappears when the tolerance threshold is not sharp anymore
\cite{Kozma:2007}.

\ack The authors are partially supported by the EU under contract
001907 (DELIS).

\end{document}